\documentclass[letterpaper, 10 pt, conference]{ieeeconf}  % Comment this line out if you need a4paper

\IEEEoverridecommandlockouts                              % This command is only needed if 
\usepackage{graphics}  % for pdf, bitmapped graphics files
\usepackage{amsmath}   % assumes amsmath package installed
\usepackage{amssymb}   % assumes amsmath package installed
\usepackage{esdiff}
\usepackage{xcolor}
\usepackage{booktabs}
\usepackage{array}
\usepackage{graphicx}
\title{\LARGE \bf
Modeling and Estimation of Solid Electrolyte Interphase\\ during Formation
in Battery Manufacturing
}

\author{Zhiwen Wan$^{1,*}$, Hamidreza Movahedi$^{1}$, Wenxue Liu$^{1}$, Jingchen Ma$^{1}$, \\ Jason B. Siegel$^{1}$, Andrew Weng$^{1,*}$ and Anna Stefanopoulou$^{1,*}$% <-this % stops a space
\thanks{$^{1}$Zhiwen Wan, Hamidreza Movahedi, Wenxue Liu, Jingchen Ma, Jason B. Siegel, Andrew Weng and Anna Stefanopoulou are with the Department of Mechanical Engineering, University of Michigan, Ann Arbor, MI, 48105 USA.
        {\tt\small {zhiwen, movahedi, wenxuel, jingchma, siegeljb, asweng, annastef}@umich.edu}}%
\thanks{$^{*}${Correspondence: \tt\small zhiwen, asweng, annastef@umich.edu}}%
}

\date{ }

\begin{document}
\maketitle
\begin{abstract}
The solid electrolyte interphase (SEI) --  a critical passivation layer that governs the longevity, safety, and efficiency of lithium-ion batteries -- is created during the last step in cell manufacturing called cell formation. Conventional cell formation protocols are largely empirical, resulting in long processing times and limited control over the SEI growth rate that influences SEI quality and lifetime performance. This paper develops a control-oriented, semi-empirical model to estimate SEI thickness growth from terminal voltage and cell expansion measurements acquired in-operando during manufacturing using low-cost micrometer-precision integrated-sensing fixture. %The model incorporates electrode thermodynamics as well as both diffusion-limited and reaction-limited SEI formation mechanisms. 
%We compare the SEI layer observability with voltage-only or voltage and expansion measurements. 
%We demonstrate the critical role of expansion in estimating SEI. 
Model parameters are calibrated against cell formation data, and an unscented Kalman filter is employed to estimate the SEI film growth. The results lay the foundation for future closed-loop control of SEI growth, enabling high-quality and more efficient formation processes.
% This document is a model and instructions for \LaTeX.
% This and the IEEEtran.cls file define the components of your paper [title, text, heads, etc.]. *CRITICAL: Do Not Use Symbols, Special Characters, Footnotes, 
% or Math in Paper Title or Abstract.
\end{abstract}

\section{Introduction}
\begin{figure} [h!]
    \centering
    \includegraphics[width=1\linewidth]{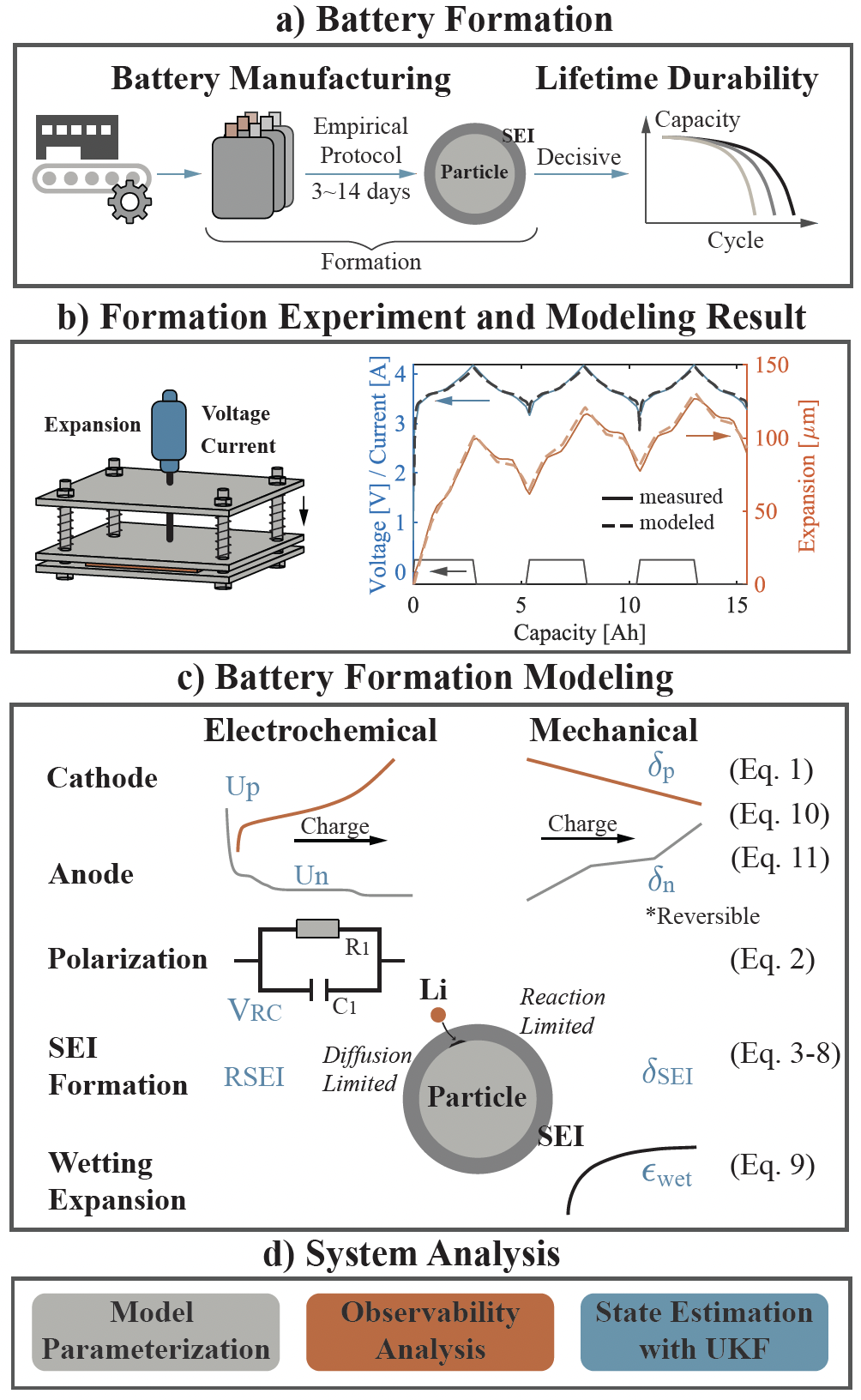}
   \caption{\textbf{Overview of the proposed framework for control-oriented modeling and estimation of SEI growth during formation.} 
\textbf{(a)} Battery formation in manufacturing: empirical protocols are time-consuming, yet decisive for lifetime performance.  
\textbf{(b)} Formation experiment and modeling result: measured current, voltage, and in-situ cell expansion, with modeled voltage and expansion.  
\textbf{(c)} Control-oriented formation model: coupling electrode expansion, SEI growth, and electrode wetting.  
\textbf{(d)} System analysis workflow: model parameterization, observability assessment, and state estimation via unscented Kalman filter (UKF).}
    \label{fig1}
\end{figure}

The solid electrolyte interphase (SEI) plays a pivotal role in the longevity and safety of lithium-ion batteries \cite{adenusi2023lithium}. The SEI forms a stable electrode–electrolyte interface that mitigates further side reactions and enables long-term cycling stability. It is primarily established during battery formation, the final stage of cell manufacturing. As shown in Fig.~1a, formation links battery manufacture to lifetime durability through the creation of the SEI, whose properties strongly influence later degradation \cite{weng2021predicting}. Yet, despite its importance, formation remains one of the most time-consuming steps in manufacturing, typically lasting 3--14 days under industrial charge--discharge protocols \cite{weng2021predicting}. These protocols are still designed largely by empirical trial-and-error, limiting our ability to actively regulate SEI growth and thereby optimize battery durability \cite{weng2021predicting}. 
% This gap motivates the present work: we aim to develop a control-oriented framework that enables systematic monitoring and regulation of SEI growth during formation, by controlling the current applied during the first charge and discharge process during manufacturing. We hypothesize that certain rates of SEI growth might produce more durable SEI than others. So, estimating the intra-cycle SEI pattern as it dynamically forms is as important as quantifying the inter-cycle SEI growth (which is currently measured with high accuracy Coulombic integration at the end of each cycle). Ultimately, this will be useful in deliberately engineering high-quality SEI and reducing formation cost and duration.

% To actively regulate the formation process and ultimately improve battery longevity, it is essential to first develop models that enable observation and estimation of SEI states during charge and discharge.
To move from empirical formation protocols toward closed-loop regulation, a control-oriented model is needed in which the states of interest can be estimated from measurable outputs. Classical control-oriented formulations describe SEI growth as either diffusion-limited~\cite{pinson2012theory} or reaction-kinetics-limited~\cite{zhou2017battery}, while more recent mixed-mode models combine both mechanisms to capture observed growth laws and thickness evolution~\cite{kamyab2019mixed}. Although these models provide valuable mechanistic insight, most efforts have focused on degradation prediction rather than the system-theoretic properties required for feedback control. In particular, a prior study \cite{zhou2017battery} has shown that, under typical operating conditions, voltage-only measurements provide limited structural identifiability and observability of SEI-related states and parameters. This limitation suggests that an additional in-situ measurement channel is needed to make SEI dynamics accessible for estimation and control.

Among candidate sensing modalities, cell expansion is especially attractive because it is non-destructive and sensitive to the irreversible thickness growth associated with SEI formation \cite{pannala2022low}. As illustrated in Fig.~1b, we therefore propose a simplified control-oriented electrochemical model with expansion as an additional state, and use voltage and expansion as joint measurements. Within the framework summarized in Fig.~1c-d, this paper makes three contributions: model parameterization from formation data, observability analysis of SEI-related states, and unscented Kalman filter (UKF)-based state estimation. We show that incorporating expansion overcomes the limited observability of voltage-only sensing and enables estimation of SEI thickness and the associated side-reaction current during formation. Our results therefore establish voltage-expansion coupled sensing as a viable pathway for future closed-loop regulation of SEI formation.
%, reducing the condition number of the observability matrix by nearly three orders of magnitude compared to voltage-only sensing. In this sense, our work resolves prior limitations and establishes that SEI growth can indeed be made observable during the formation step. An overview of the proposed approach is illustrated in Figure~\ref {fig1}. The remainder of the paper is organized as follows: Section II introduces the battery formation model, its parameterization using experimental data, and observability analysis. Section III presents the implementation of the Unscented Kalman Filter (UKF) together with practical considerations.

% The proposed model is parameterized using experimental formation data. Building on this parameterized model, we design an Unscented Kalman Filter (UKF) that exploits both voltage and expansion measurements to estimate SEI thickness in real time. The UKF framework is selected for its ability to handle nonlinear state dynamics and measurement equations without requiring explicit linearization, thereby providing a practical and computationally efficient tool for online monitoring of SEI growth during formation.

\section{Battery Model and SEI Dynamics}
This section introduces the control-oriented model used to describe cell formation, with emphasis on SEI growth dynamics and their coupling with voltage and expansion outputs. 
The model is intentionally simplified relative to physics-based formulations to enable tractable analysis of observability and state estimation. It captures the dominant electrochemical and mechanical effects during the early cycles of formation while remaining suitable for control-oriented studies. 

\subsection{Control-Oriented Model Structure}
The control-oriented model structure is summarized in Fig.~1c and consists of the following elements:
\begin{itemize}
    \item A bucket model \cite{han2014comparative} is used, with electrolyte dynamics and surface kinetics lumped into an empirical polarization branch.
    \item Average stoichiometries $x_n, x_p$ represent the lithiation states in anode and cathode active materials. Their respective open-circuit potentials are $U_n,U_p$.
    \item An RC polarization voltage $V_{RC}$ lumps anode charge-transfer and solid-state diffusion effects.
    \item SEI growth is modeled as a parasitic side reaction at the anode, with a mixed kinetic--diffusion limitation combined via Koutecký--Levich formulation.
    \item Wetting expansion $\varepsilon_{\mathrm{wet}}$ accounts for irreversible structural changes in the anode during initial lithiation, such as graphite layer rearrangement and electrolyte molecular intercalation \cite{besenhard1995filming,gao2022mechanics}. 
    % Since these effects are closely linked to electrode wetting, we lumped them together in this state.
% .
%     \item Temperature is assumed constant; thermal coupling is treated separately.
\end{itemize}

\subsection{Battery Formation Model}
% The state vector is
% \[
% x = \begin{bmatrix}
% x_n & x_p & V_{RC} & \delta_{\mathrm{SEI}} & \varepsilon_{\mathrm{wet}}
% \end{bmatrix}^\top,\quad
% u = I,
% \]
% where positive applied current $I>0$ denotes charging.

% The outputs are
% \[
% y = \begin{bmatrix}
% V_{\mathrm{cell}} \: \: \:\varepsilon_{\mathrm{cell}}
% \end{bmatrix}^\top.
% \]

\paragraph*{Electrode dynamics}
\begin{align}
\dot{x}_n &= \frac{I - j_{\mathrm{SEI}} A L_n a_{s,n}}{C_n}, &
\dot{x}_p &= -\frac{I}{C_p}.
\label{eq:electrode}
\end{align}
where positive applied current $I>0$ denotes charging.
\paragraph*{Polarization dynamics}
\begin{equation}
\dot{V}_{RC} = -\frac{1}{R_{1}C_{1}},V_{RC} + \frac{1}{C_{1}}\,I.
\label{eq:vrc}
\end{equation}

\paragraph*{SEI thickness growth dynamics}
The SEI thickness increase causes irreversible expansion \cite{wan2025degradation} on the cell level. The amount of SEI generated during formation is significantly more than that used later.
\begin{align}
\dot{\delta}_{\mathrm{SEI}} &= \frac{V_{\mathrm{m,SEI}}}{nF}\, j_{\mathrm{SEI}}, \label{eq:delta}\\
j_{\mathrm{SEI}}^{-1} &= j_{\mathrm{kin}}^{-1} + j_{\mathrm{dif}}^{-1}, \\
j_{\mathrm{kin}} &= nF k_{\mathrm{SEI}} c_{\mathrm{EC}}^0 
\exp\!\left(-k_{\mathrm{BV}} \,\eta_s\right), \\
j_{\mathrm{dif}} &= \frac{nF D_{\mathrm{SEI}} c_{\mathrm{EC}}^0}{\max(\delta_{\mathrm{SEI}},\delta_{\min})}, \\
\eta_s &= -V_{RC} + U_n(x_n) - U_{\mathrm{SEI}}\\
R_{\mathrm{SEI}} &= \frac{\delta_{\mathrm{SEI}}}{\kappa_{\mathrm{SEI}}A L_n a_{s,n}}.
\end{align}

\paragraph*{Wetting expansion dynamics}
The additional expansion term $\varepsilon_{\mathrm{wet}}$ is modeled as a first--order relaxation toward a constant steady--state value 
$\varepsilon_{\mathrm{wet}}^{\mathrm{ss}}$ with time constant $\tau_{\mathrm{wet}}$:
\begin{equation}
{\tau_{\mathrm{wet}}}\dot{\varepsilon}_{\mathrm{wet}}
= g(I)\,(\varepsilon_{\mathrm{wet}}^{\mathrm{ss}} - \varepsilon_{\mathrm{wet}}),
\end{equation}
where $g(I)$ is a step function that activates the dynamics only during charging 
($g(I)=1$ if $I>0$, and $g(I)=0$ otherwise). 

This formulation captures the gradual relaxation of structural changes during formation, most notably the wetting expansion of the graphite anode. In this process, electrolyte molecules co-insert with lithium ions into the graphene layers, leading to a larger initial expansion as described in \cite{besenhard1995filming}. The effect is irreversible and stabilizes once the formation process is completed. Since the wetting rate depends on the lithiation speed of the anode, we model the relaxation time constant $\tau_{\mathrm{wetting}}$ as current-dependent: higher charging currents shorten the effective time constant, while lower currents prolong the wetting process. To reflect its physical occurrence, the step function ensures that wetting expansion is activated only during charging. During discharge, the inserted electrolyte molecules remain within the anode structure and the wetting contribution ceases or proceeds negligibly.
\begin{table}[t]
\caption{Model parameters and units.}
\centering
\begin{tabular}{lll}
\hline
Symbol & Description & Unit \\ \hline
$C_n, C_p$ & Effective anode/cathode capacities & As \\
$R_{1}, C_{1}$ & RC branch resistance and capacitance & $\Omega$, F \\
$R_e$ & Lumped ohmic resistance & $\Omega$ \\
$\delta_{\mathrm{SEI}}$ & SEI thickness & m \\
$\kappa_{\mathrm{SEI}}$ & Ionic conductivity of SEI & S/m \\
$A$ & Electrode surface area & m$^2$ \\
$L_n$ & Anode thickness & m \\
$a_{s,n}$ & Anode specific surface area & m$^2$/m$^3$ \\
$F$ & Faraday constant & C/mol \\
$R$ & Universal gas constant & J/mol$\cdot$K \\
$T$ & Temperature & K \\
$U_{\mathrm{SEI}}$ & SEI reference potential & V \\
$k_{\mathrm{SEI}}$ & SEI reaction rate constant & m/s \\
$D_{\mathrm{SEI}}$ & Diffusion coefficient through SEI & m$^2$/s \\
$c_{\mathrm{EC}}^0$ & Bulk EC concentration & mol/m$^3$ \\
$j_{\mathrm{kin}}, j_{\mathrm{dif}}$ & Kinetic / diffusion-limited current density & A/m$^2$ \\
$V_{m,\mathrm{SEI}}$ & SEI molar volume & m$^3$/mol \\
$k_{\mathrm{BV}}$ & Butler–Volmer slope constant $(\alpha_c nF/RT)$ & 1/V \\
$\delta_{\min}$ & Minimum SEI thickness (regularization) & m \\
$U_n,U_p$ & OCP function & V \\
$\delta_n,\delta_p$ & Electrode reversible expansion function & m \\
$\varepsilon_{\mathrm{wet}}$ & Wetting expansion state & m \\
$\varepsilon^{ss}_{\mathrm{wet}}$ & Steady-state wetting expansion & m \\
$\tau_{\mathrm{wet}}$ & Wetting relaxation time constant & s \\
$g(I)$ & Current-dependent step function & -- \\
$\delta_0, \varepsilon_0$ & Initial offsets (SEI, expansion) & m \\
$\kappa_p, \kappa_n, \kappa_s$ & Weighting factors for expansion & -- \\
$\varepsilon_{\mathrm{cell}}$ & Cell expansion (output) & m \\
$V_{\mathrm{cell}}$ & Cell terminal voltage (output) & V \\
\hline
\end{tabular}
\label{table1}
\end{table}

\paragraph*{Output equations}
\begin{align}
V_{\mathrm{cell}} &= U_p(x_p) - U_n(x_n) 
+ V_{RC} + I(R_e + R_{\mathrm{SEI}}), \\
\varepsilon_{\mathrm{cell}} &= \kappa_p \delta_p(x_p) 
+ \kappa_n \delta_n(x_n) 
+ \kappa_s \delta_{\mathrm{SEI}} 
+ \varepsilon_{\mathrm{wet}} - \varepsilon_0.
\end{align}

We can present the dynamic equations (Eq. 1-9) and the output equations (Eq. 10-11) in the general nonlinear form 
\begin{gather}
\label{eq: state space}
\dot{x}=f(x,u), y=h(x,u) 
\end{gather}
with states $x=\big[x_n,\;x_p,\;V_{RC},\;\delta_{\mathrm{SEI}},\; \varepsilon_{\mathrm{wet}}\big]^\top \in \mathcal{R}^5$, $f(.): \mathcal{R}^5 \rightarrow \mathcal{R}^5$, and $h(.) :\mathcal{R}^5 \rightarrow \mathcal{R}^2$.
The measured output vector is
\begin{equation}
y=\begin{bmatrix} V_{\mathrm{cell}} \\ \varepsilon_{\mathrm{cell}} \end{bmatrix}.
\label{eq:meas}
\end{equation}

\subsection{Formation Experiment}
The model was parameterized using experimental data collected on multi-layer pouch cells during formation cycling, as described in detail by Weng \emph{et al.}~\cite{weng2023modeling}. Briefly, the formation protocol consists of three charge-discharge cycles at a rate of 0.25 A (C/10). Thickness expansion is measured in-situ with a linear displacement sensor under controlled temperature and pressure conditions, while voltage and current signals are logged simultaneously.

\subsection{Parameterization and Fit Quality}

Model parameters were calibrated against the full formation record by minimizing a joint least–squares cost,
\begin{align}
J(\theta) = \sum_{k} \Big(
   & w_V \big[V_{\mathrm{cell}}^{\mathrm{mod}}(t_k;\theta)
             - V_{\mathrm{cell}}^{\mathrm{meas}}(t_k)\big]^2 \nonumber \\
 + & w_\varepsilon \big[\varepsilon_{\mathrm{cell}}^{\mathrm{mod}}(t_k;\theta)
             - \varepsilon_{\mathrm{cell}}^{\mathrm{meas}}(t_k)\big]^2
\Big),
\label{eq:cost}
\end{align}
with weights $w_V,w_\varepsilon$ chosen so that both channels contribute comparably to the cost.
The estimated parameter set is
\[
\theta=\{\;\varepsilon^{ss}_{\mathrm{wet}},\ \tau_{\mathrm{wet}},\ D_{\mathrm{SEI}},\ k_{\mathrm{SEI}},\ C_n,\ C_p\, \ x_{n,0} \, \ x_{p,0} \;\},
\]
while all other quantities are fixed to the values reported in ~\cite{weng2023modeling}. $x_{n,0}$ and $x_{p,0}$ denote the initial stoichiometries of the anode and cathode. 
The SEI–to–stack expansion scaling for the pouch fixture is set to $\kappa_s=28$, reflecting the stack geometry (7 active anode layers with four particles per layer), i.e., a unit particle–scale SEI thickening produces $28\times$ cell–thickness change at the stack level.

\begin{figure} [t]
    \centering
    \includegraphics[width=1\linewidth]{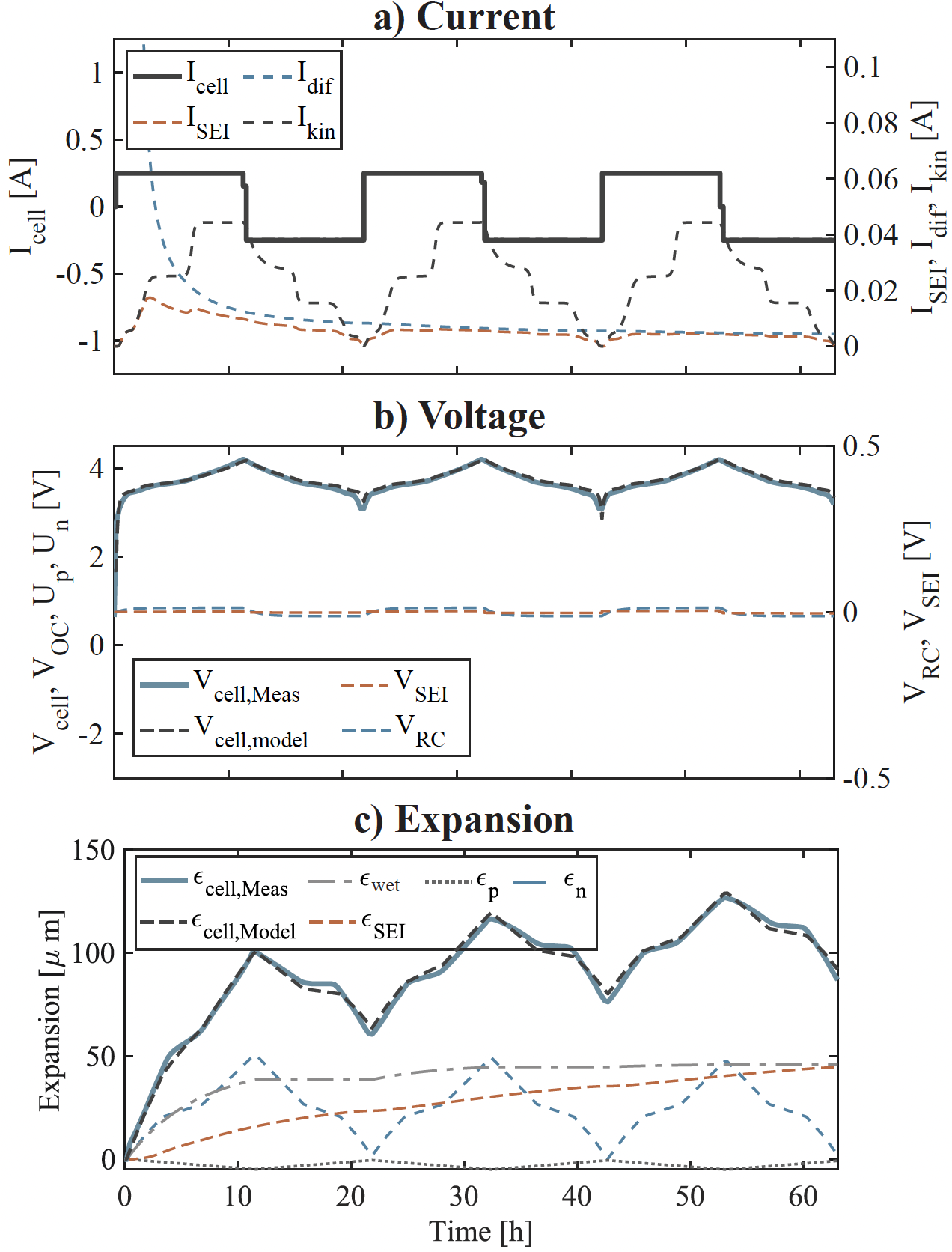}
    \caption{\textbf{Model simulation results during formation.} 
\textbf{(a)} Current decomposition shows that SEI growth is initially kinetically limited and later transitions to diffusion-limited behavior. 
\textbf{(b)} The SEI overpotential $V_{\mathrm{SEI}}$ remains unobservable from terminal voltage. 
\textbf{(c)} Expansion signals capture contributions from both the anode and SEI, providing complementary observability to voltage measurements.}

    \label{fig2}
\end{figure}

The fitted model achieves a voltage RMSE of $97.6~\mathrm{mV}$ and an expansion RMSE of $3.57~\mu\mathrm{m}$. Table \ref{tab:estimation} summarizes all the parameter fitting results. Figure~\ref{fig2} summarizes the fit for all the key quantities:

\begin{table}[t]
\centering
\caption{Estimated parameter values from formation data.}
\begin{tabular}{llll}
\hline
$\varepsilon^{ss}_{\mathrm{wet}}$ & $46.12$ &
$\tau_{\mathrm{wet}}$ & $7.7\times 10^{3}$ \\
$D_{\mathrm{SEI}}$ & $1.9\times 10^{-17}$ &
$k_{\mathrm{SEI}}$ & $9.98\times 10^{-17}$ \\
$C_n$ & $1.15\times 10^{4}$ &
$C_p$ & $1.08\times 10^{4}$ \\
$x_{n,0}$ & $8.15\times 10^{-5}$ &
$x_{p,0}$ & $1.0023$ \\
\hline
\end{tabular}
\label{tab:estimation}
\end{table}

% \begin{itemize}
% \item \textbf{Fig.~\ref{fig2}a (Current).} The SEI reaction current ($I_{\mathrm{SEI}}$) exhibits a staircase envelope governed by the anode potential and the SEI overpotential, whereas the diffusion–limited contribution ($I_{\mathrm{dif}}$) follows the expected $\sqrt{t}$–type behavior set by the growing SEI thickness. The initial charge is predominantly \emph{kinetics–limited} and transitions to \emph{diffusion–limited} except at very low SoC.
% \item \textbf{Fig.~\ref{fig2}b (Voltage).} At the low formation current ($0.025~\mathrm{A}$), polarization from the RC branch and $R_{\mathrm{SEI}}$ is small; the terminal voltage is therefore dominated by the OCPs $U_p(x_p)$ and $U_n(x_n)$. The model reproduces the measured trace with high fidelity.
% \item \textbf{Fig.~\ref{fig2}c (Expansion).} The measured thickness evolution is captured by the sum of three contributions: anode swelling $\kappa_n\delta_n(x_n)$, SEI thickening $\kappa_s\delta_{\mathrm{SEI}}$, and a throughput–modulated ``other'' term $\varepsilon_{\mathrm{wet}}$. The trend and amplitude match closely, consistent with the reported $3.57~\mu\mathrm{m}$ RMSE.
% \end{itemize}

\paragraph*{Current (Fig.~\ref{fig2}a)} 
The SEI reaction current ($I_{\mathrm{SEI}}$) develops a staircase-like envelope determined by the anode potential and the SEI overpotential, while the diffusion-limited component ($I_{\mathrm{dif}}$) follows the expected $\sqrt{t}$-type dependence associated with the growth of SEI thickness. During the initial charge, the process is primarily kinetics-limited and progressively transitions to a diffusion-limited regime, except at a very low state of charge (SoC).

\paragraph*{Voltage (Fig.~\ref{fig2}b)} 
At the applied low formation current ($0.025~\mathrm{A}$), polarization effects from the RC branch and SEI resistance are negligible. As a result, the terminal voltage is dominated by the open-circuit potentials of the cathode $U_p(x_p)$ and anode $U_n(x_n)$. The model successfully reproduces the measured voltage trace with high fidelity, confirming the accuracy of the electrochemical representation.

\paragraph*{Expansion (Fig.~\ref{fig2}c)} 
The measured cell thickness evolution is decomposed into three distinct contributions: anode swelling $\kappa_n \delta_n(x_n)$, SEI thickening $\kappa_s \delta_{\mathrm{SEI}}$, and a throughput-dependent residual term $\varepsilon_{\mathrm{wet}}$. The combined model output closely tracks both the trend and magnitude of the experimental expansion, achieving an RMSE of $3.57~\mu\mathrm{m}$.
\subsection{Simulation Results Under Higher C-rate}

\begin{figure} [t]
    \centering
    \includegraphics[width=0.975\linewidth]{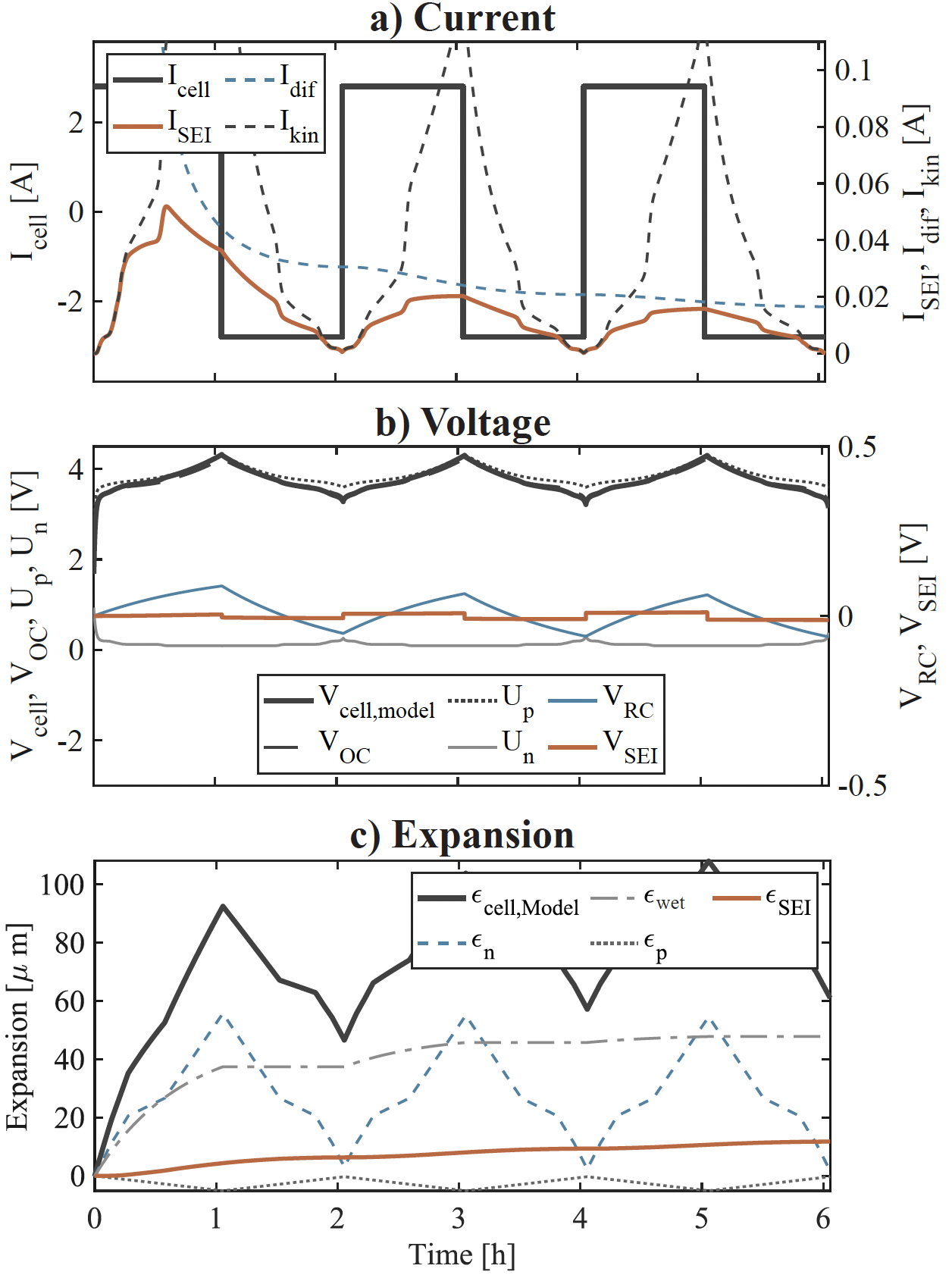}
    \caption{\textbf{Simulation of SEI growth under higher C-rate.} 
    \textbf{(a)} Cell and side-reaction currents, highlighting the increase in $I_{\mathrm{SEI}}$ and the transition between kinetic- and diffusion-limited regimes. 
    \textbf{(b)} Voltage decomposition, showing stronger polarization at high current, which drives the anode potential into the SEI reaction window earlier. 
    \textbf{(c)} Expansion contributions, comparing SEI thickening, electrode swelling, and additional expansion terms.}
    \label{fig2_1}
\end{figure}

To examine the effect of current magnitude on SEI dynamics, the calibrated model was simulated under a high current input of $2.8$~A (1C) and compared against the low-current case of $0.25$~A ($\approx$C/10). As shown in Fig.~\ref{fig2_1}, the kinetic--limited contribution increases substantially under the higher current, thereby raising the overall magnitude of the SEI side--reaction current $j_{\mathrm{SEI}}$. This behavior arises because the larger polarization shifts the anode potential into the SEI reaction regime earlier during charging. Conversely, during discharge the enhanced polarization suppresses the reaction rate, leading to reduced SEI growth in that phase. Although the total formation duration is shortened by nearly an order of magnitude, the resulting SEI thickness decreases by only a factor of three. These results demonstrate that modulating the applied current enables deliberate shaping of the $j_{\mathrm{SEI}}$ trajectory, offering a pathway to formation protocols that are both more time-efficient and yield potentially more durable SEI layers.  

\subsection{Observability Analysis}

% Utilizing Lie derivative observability analysis, it can be demonstrated that the nonlinear system described in 
% Eq. \ref{eq: state space} is observable. The inclusion of expansion measurement is essential for observability of the system, as evidenced by prior research 
% \cite{di2010lithium}, which indicates that even stoichiometry states ($x_p$
% and $x_n$) are not observable within most of the cell's operational domain. Consequently, additional assumptions are generally required to estimate these states 
% \cite{lopetegi2025electrode, allam2018interconnected}. In this context, the incorporation of expansion measurement resolves the unobservability issue, rendering all five states observable. 

% % Using the nonlinear observability framework based on Lie derivatives \cite{hermann1977nonlinear}, the observability matrix associated with Eq.~\ref{eq: state space} is found to have full rank 5 under the considered operating conditions, indicating local observability of all five states. This result relies critically on the inclusion of expansion measurement. Prior work has shown that, with voltage-only sensing, even the stoichiometry states $x_p$
% and $x_n$ are not observable over much of the cell operating domain \cite{di2010lithium}, so additional assumptions are typically required for their estimation \cite{lopetegi2025electrode, allam2018interconnected}. In contrast, augmenting voltage with expansion resolves this unobservability issue and renders the full five-state system locally observable.

Using the nonlinear observability framework based on Lie derivatives \cite{hermann1977nonlinear}, the observability matrix associated with Eq.~\ref{eq: state space} has full rank 5 under the considered operating conditions, indicating local observability of all five states. This result depends critically on the inclusion of expansion measurement. Prior work has shown that, with voltage-only sensing, even the stoichiometry states $x_p$
and $x_n$ are not observable over much of the cell operating domain \cite{di2010lithium}, requiring additional assumptions for state estimation \cite{lopetegi2025electrode, allam2018interconnected}.

\section{State Estimation}
In practice, lithium-ion cells exhibit manufacturing variability:
initial states such as anode/cathode lithiation may be unknown (e.g., due to
prelithiation), and electrode capacities can vary across batches. These
uncertainties, together with process noise, mean that open-loop model
predictions are insufficient to guide control of SEI growth during formation.
Moreover, voltage and expansion signals are noisy and limited, while SEI
dynamics are strongly nonlinear, governed by coupled kinetic–diffusion
mechanisms. To robustly infer hidden states such as $\delta_{\mathrm{SEI}}$ and
to account for uncertainty in both initial conditions and measurements,
we employ the UKF, which is well-suited for
nonlinear systems with uncertain parameters and provides real-time state
estimates with quantified confidence.

\begin{table}[t]
	\caption{Unscented Kalman Filter}
	\begin{center}
		\begin{tabular}{p{8cm}}
			%\hline
			%\textbf{Table}&\multicolumn{3}{|c|}{\textbf{Table Column Head}} \\
			%	\cline{2-4} 
			\hline
			\textbf{Sigma Point Selection(unscented transformation)}\\
			$\hat{\textbf{x}}_{k-1}^{0}=\hat{\textbf{x}}_{k-1}^{+}$\\
			$\hat{\textbf{x}}_{k-1}^i=\hat{\textbf{x}}_{k-1}^{+}+(\sqrt{n+\lambda}(\sqrt{\textbf{P}_{k-1}}))_m,  i= 1,...,n$\\
			$\hat{\textbf{x}}_{k-1}^i=\hat{\textbf{x}}_{k-1}^{+}-(\sqrt{n+\lambda}(\sqrt{\textbf{P}_{k-1}}))_m, i = n+1,...,2n$\\
			*$\sqrt{\ \ \ }$: Cholesky decomposition; $n$: state number. $m = 1,...,n$. \\Superscript with $i$ presents individual sigma points.\\
			weighting coefficients for mean and covariance:\\ $w_o^m=\frac{\lambda}{n+\lambda} ,w_o^c = \frac{\lambda}{n+\lambda} +1+\beta-\alpha^2$\\
			$w_i^m=w_i^c=\frac{1}{2(n+\lambda)}, i=1,2,...,2n$ \\
			$\lambda=\alpha^2(n+\kappa)-n$\\
			$\beta=2$: Gaussian problem. $0 \le \alpha \le 1$ is an appropriate choice for $\alpha$, where a larger $\alpha$ generates more dispersed sigma points. \\
            \hline
			\textbf{Time Update} \\
			Propagation:
			$\hat{\textbf{x}}_k^i=f(\hat{\textbf{x}}_{k-1}^{i},\textbf{u}_k)$\\
			Predicted mean:	
			$\hat{\textbf{x}}_k^-=\sum_{i=0}^{2n}w_i^m\hat{\textbf{x}}_k^i$\\
			Predicted covariance:
			$\textbf{P}_k^-=\sum_{i=0}^{2n}w_i^c(\hat{\textbf{x}}_k^i-\hat{\textbf{x}}_k^-)(\hat{\textbf{x}}_k^i-\hat{\textbf{x}}_k^-)^T+\textbf{Q}_k$\\
            \hline
			\textbf{Measurement Update}\\
			Predicted outputs: $\hat{\textbf{y}}_k^i=h(\hat{\textbf{x}}_k^i,\textbf{u}_k), i=0,1,2,...,2n$\\
			Predicted mean:$	\hat{\textbf{y}}_k=\sum_{i=0}^{2n}w_i^m\hat{\textbf{y}}_k^i$\\
			Innovation covariance: $\textbf{P}_k^h=\sum_{i=0}^{2n}w_i^c(\hat{\textbf{y}}_k^i-\hat{\textbf{y}}_k)(\hat{\textbf{y}}_k^i-\hat{\textbf{y}}_k)^T+\textbf{R}_k$\\
			Cross correlation: $\textbf{P}_k^{fh}=\sum_{i=0}^{2n}w_i^c(\hat{\textbf{x}}_k^i-\hat{\textbf{x}}_k^-)(\hat{\textbf{y}}_k^i-\hat{\textbf{y}}_k)^T$\\
			Kalman gain: $\textbf{K}_k=\textbf{P}_k^{fh}(\textbf{P}_k^h)^{-1}$\\
			State update: $\hat{\textbf{x}}_k^+=\hat{\textbf{x}}_k^-+\textbf{K}_k(\textbf{y}_k-\hat{\textbf{y}}_k)$\\
			Covariance update: $\textbf{P}_k=\textbf{P}_k^--\textbf{K}_k\textbf{P}_k^h\textbf{K}_k^T$\\
			*superscript with +: after prediction; with -: before prediction.\\
			\hline
			%	\hline
			%	\multicolumn{4}{l}{$^{\mathrm{a}}$Sample of a Table footnote.}
		\end{tabular}
		\label{tab2}
	\end{center}
\end{table}

\subsection{State–Estimation Setup}
We estimate the states 
$x=[x_n,\;x_p,\;V_{RC},\;\delta_{\mathrm{SEI}},\;\varepsilon_{\mathrm{wet}}]^\top$
using a UKF \cite{julier1997new} with the steps detailed in Table \ref{tab2}. The prior
mean is the calibrated operating point $x_0^{\text{prior}}$ and the
prior covariance is
\[
\mathbf{P}_0=\mathrm{diag}\:\!\big([0.01^2,\;0.01^2,\;0.01^2,\;0.01^2,\;0.01^2]\big),
\]
i.e., standard deviations of $0.01$ in stoichiometries (–), $10$ mV in
$V_{RC}$, and $0.01~\mu$m in $\delta_{\mathrm{SEI}}$ and
$\varepsilon_{\mathrm{wet}}$.

The continuous–time process model is discretized with RK4 under
zero-order-hold current; an internal sub-stepping factor is chosen so the
integration step is $\lesssim\!1$ s. Process noise is taken as a small,
isotropic covariance,
\[
\mathbf{Q}=10^{-16}\,\mathbf{I}_5,
\]
which represents high trust on the model.

Measurements are the terminal voltage (V) and in-situ
expansion ($\mu$m). The measurement covariance is
\[
\mathbf{R}=\mathrm{diag}\:\!\big([\,4{\times}10^{-6},\;1\,]\big)
\]
corresponding to $\sigma_V=2$ mV and $\sigma_\varepsilon=1~\mu$m.

Unscented transform parameters are
\[
\alpha=10^{-3},\qquad \beta=2,\qquad \kappa=-2,
\]
with $\lambda=\alpha^2(n+\kappa)-n$.

% We use $(\alpha,\beta,\kappa)=(10^{-3},\,2,\,-2)$, 
% a small process noise $Q$ to stabilize propagation, and a diagonal $R$ reflecting voltage and expansion noise (expansion in $\mu$m).
% Outputs are optionally scaled by a factor $s_E$ in $h(\cdot)$ to balance units (as in the observability section).
% State constraints (e.g., $x_n,x_p\in[0,1]$, $\delta_{\mathrm{SEI}}\ge 0$) are enforced by soft clipping after updates.

% \subsection{Results}
% The UKF tracks both voltage and expansion with low error over the formation cycles.
% With the two-sensor configuration, $\delta_{\mathrm{SEI}}$ is estimated consistently and the innovation sequences are zero-mean and whitened.
% We report RMSEs of $0.098\,\mathrm{V}$ (voltage) and $3.6\,\mu\mathrm{m}$ (expansion), and include $2\sigma$ state credibility bands consistent with the simulated trajectories.

\subsection{State Estimation Results}
To evaluate robustness against parameter uncertainty, we tested the observer under deliberately inaccurate initialization of the electrode stoichiometries. Specifically, both anode and cathode stoichiometries were perturbed by $20\%$ to emulate the unknown prelithiation state that often arises in practical manufacturing.

\begin{figure} [t]
    \centering
    \includegraphics[width=1\linewidth]{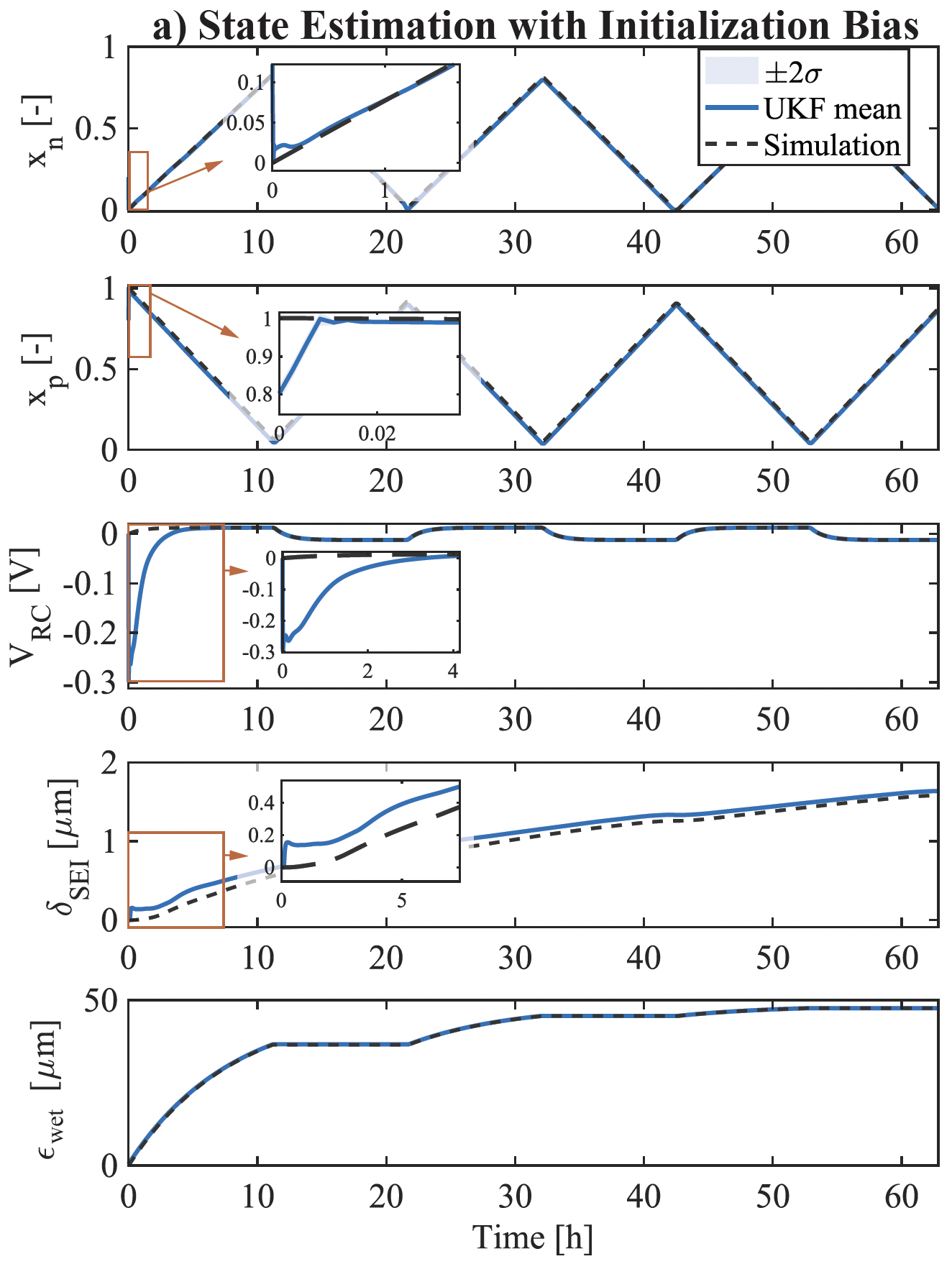}
    \caption{\textbf{State estimation with biased initialization.} 
    The UKF was initialized with a 20\% error in both anode and cathode stoichiometries to emulate uncertainty from unknown prelithiation. 
    Estimated states (solid lines) are shown together with their $\pm 2\sigma$ confidence envelopes and simulated trajectories (dashed lines) for reference. 
    Results indicate rapid correction of the anode stoichiometry within the first minute, slower convergence of the cathode stoichiometry and $V_{RC}$ over several hours, and reliable tracking of SEI thickness and wetting-related expansion despite biased initialization.}
    \label{fig4}
\end{figure}

The results demonstrate that the proposed UKF framework reliably reconstructs the coupled electrochemical and mechanical states, including SEI thickness, even under incorrect initial conditions. As shown in Fig.~\ref{fig4}, both anode and cathode stoichiometries ($x_n$, $x_p$) converge rapidly to their simulated trajectories within the first minute. This fast convergence is enabled by the steep early voltage slope and the sharp expansion response of the anode, which directly constrain the filter. The polarization state $V_{RC}$, although unbiased at initialization, exhibits a transient deviation but aligns with the simulated trajectory after about three hours.  

The SEI thickness $\delta_{\mathrm{SEI}}$ (our primary state of interest) is initially overestimated but gradually converges to the reference trajectory, confirming that the joint use of voltage and expansion measurements renders it observable. The additional expansion state $\varepsilon_{\mathrm{wet}}$, modeled as a wetting-induced exponential relaxation, closely follows the simulated trend across the entire formation process.  

\begin{figure} [t]
    \centering
    \includegraphics[width=1\linewidth]{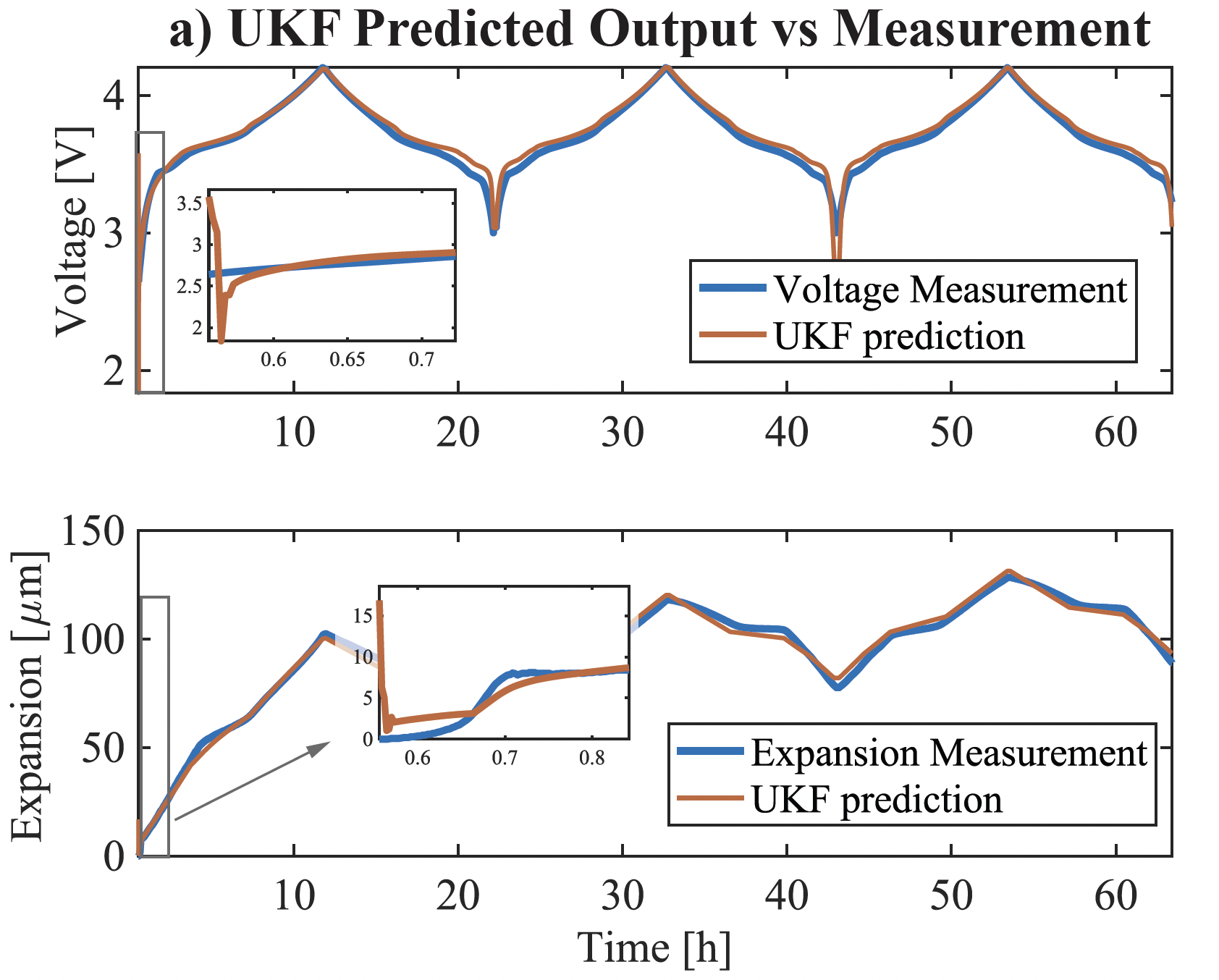}
    \caption{\textbf{UKF predicted outputs versus measurements.} 
    Comparison of UKF predictions with experimental data for terminal voltage (top) and cell expansion (bottom). 
    Once the internal states converge, the UKF reproduces both outputs with high fidelity, yielding root-mean-square errors of $55.3$~mV for voltage and $2.65~\mu$m for expansion.}
    \label{fig4_1}
\end{figure}

The agreement between predicted and measured outputs is shown in Fig.~\ref{fig4_1}. Once the internal states converge, the UKF reproduces both terminal voltage and cell expansion with high fidelity. The resulting root-mean-square errors are $55.3$~mV for voltage and $2.65~\mu$m for expansion, both lower than the open-loop model fitting error. This demonstrates that the UKF effectively balances model and measurement uncertainties to minimize residuals, thereby enhancing robustness for real-time monitoring.

\section{Conclusion and Perspective}
This work provides an exploratory first step toward understanding SEI growth during battery formation from a control perspective, and lays the foundation for model-based monitoring and future closed-loop regulation of the formation process. We proposed a control-oriented electrochemical–mechanical model and demonstrated that SEI thickness can be rendered observable by augmenting terminal voltage with in-situ expansion measurements. The model was parameterized using formation data, achieving a fitting accuracy of $97.6~\mathrm{mV}$ RMSE in voltage and $3.57~\mu\mathrm{m}$ RMSE in expansion. Implemented within a UKF, the framework enabled reliable real-time state estimation under biased initial conditions, with the key stoichiometries converging within minutes and the SEI thickness gradually aligning with the simulated trajectory. Furthermore, simulation under higher C-rates revealed that current modulation can shape the $j_{\mathrm{SEI}}$ trajectory. These results provide a foundation for observability-informed control of formation, with the potential to lower manufacturing cost and time while improving SEI quality and ultimately the long-term battery durability.

% This study has several limitations. First, the model has not yet been experimentally validated through post-mortem analysis, which would provide direct evidence for the assumed mechanisms of SEI growth and expansion. Second, parameterization and testing were performed only under low C-rate formation data, without a systematic sensitivity analysis of model parameters. Finally, while observability was demonstrated for low-current conditions, the impact of different current profiles on observability metrics, such as the condition number and the observability Gramian, remains unexplored.

This study has several limitations. First, the proposed control-oriented model relies on simplifying assumptions regarding the relationship between SEI growth and cell expansion, so the identified SEI-related states should be interpreted within the scope of the reduced-order modeling framework rather than as direct physical measurements. Second, the model has not yet been experimentally validated through post-mortem analysis, which would provide direct evidence for the assumed mechanisms of SEI growth and expansion. Third, parameterization and testing were performed only under low C-rate formation data, without a systematic sensitivity analysis of model parameters. Finally, while observability was demonstrated for low-current conditions, the impact of different current profiles on observability metrics, such as the condition number and the observability Gramian, remains unexplored.

Future efforts will focus on conducting formation experiments under different C-rates to refine model fidelity, and on integrating complementary sensing modalities such as electrochemical impedance spectroscopy to improve estimation accuracy and robustness. We also envision leveraging this framework to design and implement optimized formation current profiles, with the dual goals of reducing manufacturing cost and time while enhancing long-term battery durability.

\bibliographystyle{ieeetr}
\bibliography{main}

\end{document}